\documentclass[12pt]{article}
\begin{document}

\title{On Persistence and Stability of some Biological Systems with Cross Diffusion}
\author{\ E. Ahmed$^{1,2}$, A. S. Hegazi$^{1}$ and A. S. Elgazzar$^{3,4}$ \\
$^{1.}$Mathematics Department, Faculty of Science\\
35516 Mansoura, Egypt\\
$^{2.}$Mathematics Department, Faculty of Science\\
Al-Ain, P. O. Box 17551, UAE\\
$^{3.}$Mathematics Department, Faculty of Science - Al-Jouf\\
King Soad University, Kingdom of Saudi Arabia\\
$^{4.}$Mathematics Department, Faculty of Education\\
45111 El-Arish, Egypt}
\maketitle
\date{}

\begin{abstract}
The concept of cross diffusion is applied to some biological systems. The
conditions for persistence and Turing instability in the presence of cross
diffusion are derived. Many examples including: predator-prey, epidemics
(with and without delay), hawk-dove-retaliate and prisoner's dilemma games
are given.
\end{abstract}

\section{Introduction}

Cross diffusion is the diffusion of one type of species due to the presence
of another [Okubo 1980, Chattopadhyay and Chatterjee 2001]. This phenomena
is abundant in nature e.g. predator-prey systems where the predator diffuses
towards the regions where the prey is more abundant. On the other hand, the
prey tries to avoid predators by diffusing away from them. Another area of
application is in epidemics where susceptible individuals try to avoid
infected individuals.

Our aim is to generalize some biological systems to the case of the presence
of cross diffusion. We study the persistence and Turing instability of the
generalized models. Several examples are given e.g. predator-prey system,
epidemic model, hawk-dove-retaliate game and prisoner's dilemma game.

The paper is organized as follows: In section 2 the conditions for Turing
instability in the presence of cross diffusion are derived then applied to a
predator-prey system. Necessary conditions for persistence of a predator
prey system in the presence of cross diffusion are derived. In section 3,
cross diffusion is introduced in an epidemic model. Its effect on wave
solution with and without delay is derived. In section 4, the effect of
cross diffusion on evolutionarily stable strategy is studied.
Hawk-dove-retaliate game is given as an example. Cross diffusion is applied
to the dynamics of learning in multiagent systems in section 5. Some
conclusions are summarized in section 6.

\section{Cross diffusion in a predator-prey model}

Cross diffusion is expected to be relevant to predator-prey systems where
the predator diffuses towards the regions where the prey is more abundant.
On the other hand, the prey tries to avoid predators by diffusing away from
them. Cross diffusion term expresses the population flux of one species due
to the presence of another species. The value of the cross diffusion may be
positive, negative or zero.

Positive cross diffusion means that one type of species tends to move in the
direction of lower density of the other type and vice versa. Hence the cross
diffusion term is positive for the prey and negative for the predator.
Generalizing Lotka-Volterra predator-prey model (with prey saturation) to
include cross diffusion, one gets 
\begin{equation}
\begin{array}{l}
\frac{\partial u_{1}}{\partial t}=u_{1}(\alpha _{1}-\beta
_{1}u_{1}-u_{2})+D_{1}\frac{\partial ^{2}u_{1}}{\partial x^{2}}+\frac{%
\partial }{\partial x}(D_{12}^{\prime }(u_{1})\frac{\partial u_{2}}{\partial
x}), \\ 
\frac{\partial u_{2}}{\partial t}=u_{2}(-\alpha _{2}+u_{1})+D_{2}\frac{%
\partial ^{2}u_{2}}{\partial x^{2}}-\frac{\partial }{\partial x}%
(D_{21}^{\prime }(u_{2})\frac{\partial u_{1}}{\partial x}),
\end{array}
\end{equation}
where $D_{12}^{\prime }(u_{1})$ is the density dependent diffusion
coefficient of cross diffusion such that $D_{12}^{\prime }(u_{1})\rightarrow
0\;as\;u_{1}\rightarrow 0.\;$Following Chattopadhyay and Chatterjee
[Chattopadhyay and Chatterjee 2001] we assume that 
\[
D_{12}^{\prime }=D_{12}u_{1}/(\epsilon +u_{1})\approx D_{12}(1-\epsilon
/u_{1}+...), 
\]
where $\epsilon $ is very small hence 
\[
D_{12}^{\prime }\approx D_{12}\;\;\forall u_{1}>>\epsilon . 
\]

Similarly for $D_{21}$ thus we get

\begin{equation}
\begin{array}{l}
\frac{\partial u_{1}}{\partial t}=u_{1}(\alpha _{1}-\beta
_{1}u_{1}-u_{2})+D_{1}\frac{\partial ^{2}u_{1}}{\partial x^{2}}+D_{12}\frac{%
\partial ^{2}u_{2}}{\partial x^{2}}, \\ 
\frac{\partial u_{2}}{\partial t}=u_{2}(-\alpha _{2}+u_{1})+D_{2}\frac{%
\partial ^{2}u_{2}}{\partial x^{2}}-D_{21}\frac{\partial ^{2}u_{1}}{\partial
x^{2}},
\end{array}
\end{equation}
where $u_{1}(u_{2})$ is the prey (predator) density, and all the constants $%
\alpha _{1},\;\alpha _{2},\;\beta _{1},\newline
D_{1},\;D_{2},\;D_{12},\;D_{21}$ are nonnegative. We have used $-D_{21}$
since predators move to regions with higher prey density. The value $\alpha
_{1}$ is the net growth rate of the prey population in the absence of
predators, while $\alpha _{2}$ is the net death rate of predators due to the
absence of prey. The term $\beta _{1}u_{1}$ corresponds to the competition
within the prey species.

The existence of solutions of systems similar to (2) has been studied in
[Chen et al 2003 and Le 2003].

This system has a unique homogeneous coexistence solution 
\begin{equation}
u_{1}^{*}=\alpha _{2},\;\;\;u_{2}^{*}=\alpha _{1}-\alpha _{2}\beta
_{1},\;\;\;\;\alpha _{1}-\alpha _{2}\beta _{1}>0.\;\;
\end{equation}
It is straightforward to see that the solution (3) is asymptotically stable
in the case of no diffusion $(D_{1}=0,\;D_{2}=0,\;D_{21}=0,\;D_{12}=0).$

In Turing instability [Okubo 1980], it is required to see whether diffusion
terms can destabilize the steady state (2). We derive Turing instability in
the presence of cross diffusion: Consider two interacting species with
different diffusion coefficients 
\begin{eqnarray}
\frac{\partial u_{1}}{\partial t} &=&f_{1}(u_{1},u_{2})+D_{1}\frac{\partial
^{2}u_{1}}{\partial x^{2}}+D_{12}\frac{\partial ^{2}u_{2}}{\partial x^{2}},
\\
\frac{\partial u_{2}}{\partial t} &=&f_{2}(u_{1},u_{2})+D_{2}\frac{\partial
^{2}u_{2}}{\partial x^{2}}+D_{21}\frac{\partial ^{2}u_{1}}{\partial x^{2}}. 
\nonumber
\end{eqnarray}
Assume that $(u_{1}^{*},u_{2}^{*})$ are the spatially uniform steady states
i.e. $f_{j}(u_{1}^{*},u_{2}^{*})=0,\;j=1,2$. To examine the linear stability
of $(u_{1}^{*},u_{2}^{*})$ let 
\begin{equation}
u_{j}=u_{j}^{*}+\varepsilon _{j}\exp (ikx+\lambda t),\;\;\;\;j=1,2.
\end{equation}
One gets

\[
\left| 
\begin{array}{ll}
\lambda +D_{1}k^{2}-a_{11} & -a_{12}+D_{12}k^{2} \\ 
-a_{21}+D_{21}k^{2} & \lambda +D_{2}k^{2}-a_{22}
\end{array}
\right| =0, 
\]
where

\begin{eqnarray*}
\lambda &=&\frac{1}{2}\left[ (b_{11}+b_{22})\pm \sqrt{%
(b_{11}+b_{22})^{2}-4(b_{11}b_{22}-b_{12}b_{21})}\right] , \\
b_{jj} &=&a_{jj}-D_{j}k^{2},\;\;\;b_{ij}=a_{ij}-D_{ij}k^{2},\;i\neq j, \\
a_{ij} &=&\frac{\partial f_{i}(u_{1}^{*},u_{2}^{*})}{\partial u_{j}}.
\end{eqnarray*}
Turing instability occurs when $(u_{1}^{*},u_{2}^{*})$ is asymptotically
stable if diffusion does not exist $(k=0)$; but unstable if diffusion exists 
$(k\neq 0)$. Thus the conditions for Turing instability are 
\[
a_{11}+a_{22}<0,\;\;a_{11}a_{22}-a_{12}a_{21}>0,%
\;b_{11}b_{22}-b_{12}b_{21}<0. 
\]
The third condition implies

\[
\begin{array}{l}
(D_{1}D_{2}-D_{12}D_{21})k^{4}-(D_{1}a_{22}+D_{2}a_{11}-D_{12}a_{21}-D_{21}a_{12})k^{2}+
\\ 
(a_{11}a_{22}-a_{12}a_{21})<0,\;\;\;\;\forall k.
\end{array}
\]
Minimizing the left hand side of the above inequality with respect to $k$
one gets the conditions for Turing instability in the following form 
\begin{equation}
(D_{1}a_{22}+D_{2}a_{11}-D_{12}a_{21}-D_{21}a_{12})>2\sqrt{%
(a_{11}a_{22}-a_{12}a_{21})(D_{1}D_{2}-D_{12}D_{21})}>0.
\end{equation}

Now we consider Turing instability for the system (2). Following the above
procedure one gets:\newline
\newline
\textbf{Proposition 1:} The system (2) is Turing stable if 
\begin{equation}
D_{21}>0,\;\;D_{12}>0
\end{equation}
\newline
\textbf{Proof}: For the system (2), we have 
\[
a_{11}=-\beta _{1}\alpha _{2}<0,\;\;a_{12}=-\alpha
_{2}<0,\;\;\;a_{21}=\alpha _{1}-\beta _{1}\alpha _{2}>0,\;\;\;a_{22}=0.\; 
\]
Since $D_{1}\geq 0,\;D_{2}\geq 0,$ then 
\[
(D_{1}a_{22}+D_{2}a_{11}-D_{12}a_{21}-D_{21}a_{12})<0, 
\]
hence the condition (6) is not satisfied.

Now the persistence of the system (2) is discussed:\newline
\newline
\textbf{Definition 1:} A partial differential equation 
\[
F(t,x,u(t,x),\partial u/\partial t,\partial u/\partial x,...)=0, 
\]
defined on a spatial domain $\Omega $ is persistent in the context of zero
Dirichlet boundary conditions if there exist a region $V=\{u:\;0<u<w\}$ in $%
\Omega $ such that all solutions with nontrivial, nonnegative initial
conditions are attracted to $V$.\newline
\newline
Equivalently the boundary of the positive cone in the space where the
solutions exist act as a repeller.

Persistence [Hofbauer and Sigmund 1998] is an important concept in
population biology since it means that a certain species will continue to
exist. The work of Cantrell et al [Cantrell et al 1993] is basic in finding
sufficient conditions for persistence in some partial differential equations
without cross diffusion. Therefore we will use it to derive the necessary
conditions for persistence of the system (2). Here zero Dirichlit boundary
conditions are assumed:

\[
u_{i}=0\;\mathrm{at}\;x=0,\;x=L. 
\]

\vspace{0pt}For the sake of completeness we mention the following results
[Cantrell et al 1993] with slight modifications:\newline
\newline
\textbf{Theorem (1)}:

\begin{enumerate}
\item[(i)]  Suppose that $f(\overrightarrow{r},u)$\ is Lipschitz in $%
\overrightarrow{r}$ and continuously differentiable in $u$ with
\end{enumerate}

\[
\partial f/\partial u\leq 0\;\mathrm{for}\;u\geq 0,\;f(\overrightarrow{r}%
,u)\leq 0\;\mathrm{if}\;u\geq l, 
\]

\begin{enumerate}
\item[ ]  for some constant $l$, and $f(\overrightarrow{r},0)>0$ at some
point in the domain $\Omega $ , then the problem 
\begin{equation}
\partial u/\partial t=D\nabla ^{2}u\;+uf(\overrightarrow{r}%
,u)\;\;\;\;\;\;in\;\Omega \times (0.\infty ),
\end{equation}

\item[ ]  with Dirichlet or Neumann boundary conditions has a unique
positive steady state $u^{ss}$ which is a global attractor for nontrivial
non-negative solutions (hence the system (8) is persistent) if the following
problem has a positive eigenvalue $\sigma $%
\[
\sigma u=D\nabla ^{2}u\;+uf(\overrightarrow{r},0)\;\;\;\;\;\;in\;\Omega \;,
\]
with the same boundary conditions as (8).

\item[(ii)]  Suppose that $f_{1}(\overrightarrow{r},u_{1},u_{2})$\ and\ $%
f_{2}(\overrightarrow{r},u_{1},u_{2})$ are $C^{2}$,bounded, that $f_{1}(%
\overrightarrow{r},u_{1},0)$ satisfies the conditions of part (i) with
positive $\sigma $, that $f_{2}(\overrightarrow{r},0,u_{2})\leq 0$ for $%
u_{2}\geq 0.$ Let $\sigma _{0}$ be the largest eigenvalue of the system 
\[
\sigma _{0}u_{1}=D_{1}\nabla ^{2}u_{1}\;+u_{1}f_{1}(\overrightarrow{r}%
,0,0)\;\;\;\;\;\;in\;\Omega ,
\]
with Dirichlet or Neumann boundary conditions. Also let $\sigma _{2}$be the
largest eigenvalue of the system 
\[
\sigma _{2}u_{2}=D_{2}\nabla ^{2}u_{2}\;+u_{2}f_{2}(\overrightarrow{r}%
,u_{1}^{ss},0)\;\;\;\;\;\;in\;\Omega ,
\]
where $u_{1}^{ss}$ is given by $f_{1}(\overrightarrow{r},u_{1}^{ss},0)=0,$
then the following system is persistent if both $\sigma _{0}$\ and\ $\sigma
_{2}$ are \ positive 
\begin{eqnarray}
\partial u_{1}/\partial t &=&D_{1}\nabla ^{2}u_{1}\;+u_{1}f_{1}(%
\overrightarrow{r},u_{1},u_{2}), \\
\;\partial u_{2}/\partial t &=&D_{2}\nabla ^{2}u_{2}\;+u_{2}f_{2}(%
\overrightarrow{r},u_{1},u_{2})\;\ \ \ \ in\;\Omega \times (0.\infty ). 
\nonumber
\end{eqnarray}
\end{enumerate}

Now we consider the persistence of system (2). The first persistence
criteria is that the prey model alone (without predators) should\ be
persistent, hence [Cantrell et al 1993]

\begin{equation}
\alpha _{1}>D_{1}\left( \frac{\pi }{L}\right) ^{2}.
\end{equation}
The second criteria is that the predator system should be persistent when
the prey is at its maximum capacity $u_{1}=\alpha _{1}/\beta _{1}$, hence

\begin{equation}
\frac{\alpha _{1}}{\beta _{1}}-\alpha _{2}>D_{2}\left( \frac{\pi }{L}\right)
^{2}.
\end{equation}

Now we study persistence in the presence of cross diffusion. The most
critical region is where both $u_{1}$ and\ $u_{2}$ are small, hence we
assume that

\[
u_{j}=\varepsilon _{j}\exp (\lambda t)\sin \left( \frac{\pi x}{L}\right)
,\;\;\;\;j=1,2, 
\]
and linearize in $\varepsilon _{j},$ we get the following sufficient
conditions for persistence $(\lambda >0).$\newline
\newline
\textbf{Proposition 2: }In addition to Eqs. (9) and (10) one of the
following conditions is necessary for the persistence of the system (2):

\begin{equation}
\alpha _{1}-\alpha _{2}>\left( D_{1}+D_{2}\right) \left( \frac{\pi }{L}%
\right) ^{2},
\end{equation}

or

\begin{equation}
\left[ \alpha _{1}-D_{1}\left( \frac{\pi }{L}\right) ^{2}\right] \left[
\alpha _{2}+D_{2}\left( \frac{\pi }{L}\right) ^{2}\right]
>D_{12}D_{21}\left( \frac{\pi }{L}\right) ^{4}.
\end{equation}
\newline
\textbf{Proof: }Substituting \textbf{\ }$u_{j}$ in (2) and linearizing in $%
\varepsilon _{j}$ one gets 
\begin{eqnarray*}
\varepsilon _{1}[\lambda -\alpha _{1}+D_{1}(\pi /L)^{2}]+\;\varepsilon
_{2}D_{12}(\pi /L)^{2} &=&0 \\
\varepsilon _{2}[\lambda +\alpha _{2}+D_{2}(\pi /L)^{2}]-\;\varepsilon
_{1}D_{21}(\pi /L)^{2} &=&0
\end{eqnarray*}
To get a nonzero solution the determinant of the coefficients has to be zero
which gives the following equation

\[
\begin{array}{l}
\lambda ^{2}+\lambda [-(\alpha _{1}-\alpha _{2})+\left( D_{1}+D_{2}\right)
\left( \frac{\pi }{L}\right) ^{2}]+[(-\alpha _{1}+D_{1}\left( \frac{\pi }{L}%
\right) ^{2})(\alpha _{2}+D_{2}\left( \frac{\pi }{L}\right) ^{2})+ \\ 
D_{12}D_{21}\left( \frac{\pi }{L}\right) ^{4}=0.
\end{array}
\]
\textbf{\newline
}Persistence requires that the real part of one of the roots, at least,
should be positive hence

\ 
\[
\alpha _{1}-\alpha _{2}>\left( D_{1}+D_{2}\right) \left( \frac{\pi }{L}%
\right) ^{2}, 
\]

or

\[
\left[ \alpha _{1}-D_{1}\left( \frac{\pi }{L}\right) ^{2}\right] \left[
\alpha _{2}+D_{2}\left( \frac{\pi }{L}\right) ^{2}\right]
>D_{12}D_{21}\left( \frac{\pi }{L}\right) ^{4}. 
\]
This completes the proof.

\section{\textbf{Cross diffusion in epidemics}}

In this section, we will study an epidemic model where the population is
classified into susceptible (S), and infected and infective (I), according
to the health of each individual. The transition between the two states
occurs according to the following rules:

\begin{enumerate}
\item[ ]  Susceptible individual having at least one infected nearest
neighbor is infected in the next time step.

\item[ ]  Infected individuals are recovered and become susceptible with a
rate $\lambda $.
\end{enumerate}

Cross diffusion will be included, where S-individuals diffuse away from
I-individuals. This system will be studied both without and with delay. In
the first case we have

\begin{equation}
\begin{array}{l}
\frac{\partial S}{\partial t}=-SI, \\ 
\frac{\partial I}{\partial t}=SI-\lambda I-D\frac{\partial ^{2}S}{\partial
x^{2}},
\end{array}
\end{equation}
where both $\lambda $\ and\ $D$ are nonnegative constants.

Here we consider that only susceptibles are capable of diffusing. This
corresponds to the case where the disease has symptoms hence susceptibles
try to avoid infectives.

Studying the wave solution of Eq. (14)

\[
S=S(z),\;I=I(z),\;z=x-ct, 
\]
with the boundary conditions

\begin{equation}
S(\infty )\simeq 1,0\leq S(-\infty )\leq 1,\;I(\infty )=I(-\infty )\simeq 0,
\end{equation}
one gets

\[
I=c\frac{S^{\prime }}{S},\;-cI^{\prime }=-DS^{\prime \prime }-\lambda
I+SI,\;\;S^{\prime }=\frac{\mathrm{d}S}{\mathrm{d}z},\;I^{\prime }=\frac{%
\mathrm{d}I}{\mathrm{d}z}. 
\]
Substituting from the first equation and integrating the second one, we get

\[
F-cI=-D\frac{SI}{c}-\lambda c\ln S+cS. 
\]
To determine the constant ($F$), consider $z\rightarrow \infty $ i.e. the
past then Eq. (15) implies that $F=c$. Now consider the future $z\rightarrow
-\infty $, and let $S(t=\infty )=\sigma $, then the fraction of survivors $%
\sigma $ is given by the transcendental equation as follows:

\begin{equation}
\frac{1-\sigma }{\ln (\sigma )}=\lambda ,
\end{equation}
which is identical to an equation obtained by Murray [Murray 1993] in
another context.

It is straightforward to see that the homogeneous solution $(S=1,\;I=0)$ is
asymptotically stable if

\begin{equation}
\lambda <1,\;\ c\geq \sqrt{D}.
\end{equation}

Introducing delay, the system (14) becomes

\begin{equation}
\begin{array}{l}
\frac{\partial S}{\partial t}=-SI, \\ 
\frac{\partial I}{\partial t}=S(x,t-T)\;I(x,t-T)-\lambda I-D\frac{\partial
^{2}S}{\partial x^{2}}.
\end{array}
\end{equation}
This model is similar but not identical to that of Mendez [Mendez 1998].
Consider the wave solution of Eq. (18) with the boundary conditions (15),
and studying the stability of the homogeneous solution $(S=1,\;I=0)$, then
one gets

\[
\left( c-\frac{D}{c}\right) I^{\prime }-\lambda I+\sum_{n=0}\frac{(cT)^{n}}{%
n!}\frac{\mathrm{d}^{n}I}{\mathrm{d}z^{n}}\;=0. 
\]
Substituting with $I=e^{mz}$, then

\begin{equation}
\left( c-\frac{D}{c}\right) m-\lambda +\exp (cTm)=0.
\end{equation}
Denoting the left hand side of Eq. (19) by $f(m)$, then it is easy to see
that $f^{\prime }(m)>0\;\forall \;c\geq \sqrt{D}$ i.e. $f(m)$ is
monotonically increasing. Furthermore $f(0)>0\;\forall \;0<\lambda <1.$ Thus
we have\newline
\newline
\textbf{Proposition 3: }The solution $(S=1,\;I=0)$ for Eq. (18) subject to
the boundary conditions (15) is asymptotically stable for $(z\rightarrow
\infty )$, if the conditions (17) are satisfied.\newline
\newline
The derivation of the fraction of survivals $\sigma $ after the epidemic
wave is similar to the case without delay and is given by Eq. (16).

Also, we will discuss one of the mechanisms of cross diffusion namely
recruitment through signals [Saffre and Deneubourg 2002]. In this case
swarming individuals e.g. insects recruit others via signals e.g. pheromone
for ant colonies. Here we generalize the model of Saffre and Deneubourg to
include diffusion, and study the boundedness and persistence of the
solutions.

Let $u$ be the population density and $v$ be the pheromone density, then

\begin{equation}
\begin{array}{c}
\frac{\partial u}{\partial t}=(\alpha +\mu u)u(1-u)-\theta uv+D_{1}\bigskip 
\frac{\partial ^{2}u}{\partial x^{2}}, \\ 
\frac{\partial v}{\partial t}=\theta uv-v+D_{2}\bigskip \frac{\partial ^{2}v%
}{\partial x^{2}},
\end{array}
\end{equation}
where the parameters $\alpha ,\beta ,\theta ,\mu $ are nonnegative. There
are three homogeneous equilibrium solutions 
\[
(u,v)=(0,0),\;(1,0),\;\left( \frac{1}{\theta },\;\frac{1}{\theta ^{3}(\theta
-1)(\alpha \theta +\mu )}\right) . 
\]
The solution $(0,0)$ is unstable, the solution $(1,0)$ is asymptotically
stable if $\theta <1.$ The third (positive) solution is asymptotically
stable if 
\[
\theta >1,\;\mu (1-2.\theta )<\alpha . 
\]
The results of Cantrell et al [Cantrell et al 1993] are directly applicable
to the system (20), hence the solutions of Eq. (20) are persistent if 
\begin{equation}
\alpha >D_{1}\left( \frac{\pi }{L}\right) ^{2},\;\;\;\theta >1+D_{2}\left( 
\frac{\pi }{L}\right) ^{2}.
\end{equation}

To study the boundedness of the solutions to Eq. (20), we will consider only
solutions satisfying the following condition 
\[
u(0)\geq 0,\;v(0)\geq 0\Rightarrow u(t)\geq 0,\;v(t)\geq 0\;\forall t>0. 
\]
It is direct to see that $\partial u/\partial t<0\;\forall \;u>1$, hence $%
\sup \;u(t)\leq 1$. Consequently $\partial v/\partial t\leq (\theta
-1)v\;\Rightarrow \;\forall \;\theta \leq 1$, then $v(t)\leq v(0).$ Thus we
have:\newline
\textbf{Proposition 4: }

\begin{enumerate}
\item[(i)]  For $0\leq \theta \leq 1$, the solutions of the system (20) with 
$D_{1}=D_{2}=0\;$are bounded in $R_{+}^{2}=\left\{ (x,y)\;\mathrm{such\;that}%
\;x\geq 0,\;y\geq 0\right\} $.

\item[(ii)]  The solutions of Eq. (20) are persistent given that the
conditions (21) are valid.\newline
\end{enumerate}

\section{Cross diffusion and evolutionarily stable strategy}

Evolutionarily stable strategy ESS [Hofbauer and Sigmund 1998] is an
important concept in population dynamics. Consider a population in which
each individual adopts one of $k$ possible strategies. A strategy $I^{*}$ is
ESS if for all $I\neq I^{*}$ (yet close to it), then

\begin{enumerate}
\item[(i)]  $A(I^{*},I^{*})>A(I,I^{*})$; or

\item[(ii)]  If $A(I^{*},I^{*})=A(I,I^{*})$, then $A(I^{*},I)>A(I,I)$,
\end{enumerate}

where $A$ is the payoff matrix. The dynamics of ESS is given by replicator
dynamics 
\begin{equation}
\frac{\mathrm{d}p_{i}}{\mathrm{d}t}=p_{i}\left[ (Ap)_{i}-pAp\right]
,\;i=1,2,3,...,k,
\end{equation}
where $p_{i}$ is the fraction of the population adopting the strategy $i$.

Recently Hofbauer has introduced the concept of spatial game [Hofbauer
1999], where the system is given by 
\begin{equation}
\frac{\partial p_{i}}{\partial t}=p_{i}\left[ (Ap)_{i}-pAp\right] +D_{i}%
\frac{\partial ^{2}p_{i}}{\partial x^{2}}.
\end{equation}
This concept is generalized to asymmetric games [Ahmed et al. 2002]. Here we
generalize it further to contain cross diffusion in addition to ordinary
diffusion. The equations become 
\begin{equation}
\frac{\partial p_{i}}{\partial t}=p_{i}\left[ (Ap)_{i}-pAp\right] +D_{i}%
\frac{\partial ^{2}p_{i}}{\partial x^{2}}+\sum_{j\neq i}^{k}D_{ij}\frac{%
\partial ^{2}p_{j}}{\partial x^{2}}.
\end{equation}
The condition $\sum_{i}p_{i}=1$ implies\ $D_{i}-D_{k}+\sum_{j\neq
i}^{k}D_{ij}=0.$

Applying to the hawk(H)-dove(D)- retaliate(R) game [Maynard Smith 1982]
where players either fight for a given prize with value $v$ or agree to
share it. The first strategy is called hawk (H) and its drawback is that
fight costs value $c>v$. The second strategy is called dove (D) and its
drawback is that it loses against hawks. The third strategy is called
retaliate where the player imitates the strategy of the opponent. Hence he
(or she) is a hawk (dove) if the opponent is a hawk (dove). Thus the payoff
matrix $A$ is 
\[
A=\left[ 
\begin{array}{ccc}
\frac{1}{2}(v-c) & v & \frac{1}{2}(v-c) \\ 
0 & \frac{v}{2} & \frac{v}{2} \\ 
\frac{1}{2}(v-c) & \frac{v}{2} & \frac{v}{2}
\end{array}
\right] ,
\]
where $c$ and $v$ are constants and $c>v>0$. Let $p_{1}(p_{2})$ be the
fraction of population adopting H(D) strategy, respectively, then the system
(24) becomes 
\begin{eqnarray}
\frac{\partial p_{1}}{\partial t} &=&p_{1}\left[ -\frac{c}{2}+\frac{p_{1}}{%
2\left( c-\frac{v}{2}\right) }+\frac{p_{2}(v+c)}{2}-\frac{p_{1}^{2}c}{2}%
-cp_{1}p_{2}\right] + \\
&&D_{1}\frac{\partial ^{2}p_{1}}{\partial x^{2}}+D_{12}\frac{\partial
^{2}p_{2}}{\partial x^{2}},  \nonumber \\
\frac{\partial p_{2}}{\partial t} &=&p_{2}\left[ p_{1}\left( c-\frac{v}{2}%
\right) -\frac{p_{1}^{2}c}{2}-cp_{1}p_{2}\right] +D_{2}\frac{\partial
^{2}p_{2}}{\partial x^{2}}+D_{21}\frac{\partial ^{2}p_{1}}{\partial x^{2}}. 
\nonumber
\end{eqnarray}
Assuming $1\geq x\geq 0$, then linear stability analysis shows that $%
p_{1}=p_{2}=0$ is unstable if 
\begin{equation}
D_{12}D_{21}>D_{2}\left( D_{1}+\frac{c}{2\pi ^{2}}\right) .
\end{equation}
Thus cross diffusion can destabilize equilibrium solution, in some cases, if 
$D_{12},D_{21}$ are large enough.

\section{Cross diffusion and the dynamics of learning in multiagent systems}

Recently [Sato and Crutchfield 2002] have studied the dynamics of learning
in multiagent systems, where the agents use reinforcement learning. They
showed that, although the agents are not directly interacting with each
other, a collective game between them arises through their interaction with
the environment. Such interactions can be modelled via replicator type
equations.

Consider two agents $u$ and $v$ with $n$ possible actions, let $%
u_{j}(t)\;(v_{j}(t))$ be a measure of the probability that the agent $u(v)$
will take the action $j$ at time $t$. Sato and Crutchfield [Sato and
Crutchfield 2002] have shown that $u_{j}(t),v_{j}(t)$ satisfy the following
equations 
\begin{eqnarray}
\frac{\mathrm{d}u_{i}}{\mathrm{d}t} &=&u_{i}\beta _{1}\left[
(Av)_{i}-uAv\right] +\alpha _{1}u_{i}\sum_{j}u_{j}\ln \left( \frac{u_{j}}{%
u_{i}}\right) ,  \nonumber \\
\frac{\mathrm{d}v_{i}}{\mathrm{d}t} &=&v_{i}\beta _{2}\left[
(Bu)_{i}-vBu\right] +\alpha _{2}v_{i}\sum_{j}v_{j}\ln \left( \frac{v_{j}}{%
v_{i}}\right) ,
\end{eqnarray}
where $A$ and $B$ are payoff (reward) matrices for $u$ and $v$,
respectively, and $\alpha _{1},\alpha _{2},\beta _{1},\beta _{2}$ are
nonnegative constants. If $\alpha _{1}=\alpha _{2}=0$,$\;$then one regains
the standard asymmetric replicator equations [Hofbauer and Sigmund 1998,
Ahmed et al. 2002]. For simplicity we consider the following payoff matrices 
$A=\mathrm{diag}(a_{1},a_{2}),\;B=\mathrm{diag}(b_{1},b_{2}),\;a_{i}\geq
0,\;b_{i}\geq 0$, hence the system (27) becomes $(u_{1}=u,\;u_{2}=1-u,%
\;v_{1}=v,\;v_{2}=1-v)$%
\begin{eqnarray}
\frac{\mathrm{d}u}{\mathrm{d}t} &=&\beta _{1}u(1-v)\left(
u(a_{1}+a_{2})-a_{2}\right) ,  \nonumber \\
\frac{\mathrm{d}v}{\mathrm{d}t} &=&\beta _{2}v(1-u)\left(
v(b_{1}+b_{2})-b_{2}\right) .
\end{eqnarray}
There are four equilibria $(u,v)$: $E_{0}=(0,0),\;E_{1}=\left(
a_{2}/(a_{1}+a_{2}),0\right) ,\;E_{2}=\left( 0,b_{2}/(b_{1}+b_{2})\right)
,\;E_{3}=(1,1)$. The solution $E_{0}$ is asymptotically stable while $%
E_{1},E_{2},E_{3}$ are saddle points, hence\newline
\newline
\textbf{Corollary 1:} The system (28) is not persistent.\newline

Including cross diffusion the system becomes

\begin{eqnarray}
\frac{\mathrm{d}u}{\mathrm{d}t} &=&\beta _{1}u(1-v)\left(
u(a_{1}+a_{2})-a_{2}\right) +D_{12}\frac{\partial ^{2}v}{\partial x^{2}}, 
\nonumber \\
\;\frac{\mathrm{d}v}{\mathrm{d}t} &=&\beta _{2}v(1-u)\left(
v(b_{1}+b_{2})-b_{2}\right) +D_{21}\frac{\partial ^{2}u}{\partial x^{2}},
\end{eqnarray}
where $1\geq x\geq 0$, and the following Dirichlet boundary conditions are
assumed $u(0,t)=u(1,t)=v(0,1)=v(1,t)=0$. Assuming 
\[
u=u_{eq}+u_{0}\exp (\sigma t)\sin \pi x,\;v=v_{eq}+v_{0}\exp (\sigma t)\sin
\pi x,\; 
\]
where $(u_{eq},v_{eq})$ is one of the equilibrium solutions $%
E_{1},E_{2},E_{0}$, one gets\newline
\newline
\textbf{Proposition 5: }The following condition is necessary for persistence
of the system (29): 
\begin{equation}
D_{12}D_{21}>\max \left\{ \frac{\beta _{1}\beta _{2}a_{2}b_{2}}{\pi ^{4}},\;%
\frac{\beta _{1}\beta _{2}a_{1}b_{2}}{\pi ^{4}(a_{1}+a_{2})},\;\frac{\beta
_{1}\beta _{2}b_{1}a_{2}}{\pi ^{4}(b_{1}+b_{2})}\right\} .
\end{equation}
\newline
\newline
\textbf{Proof. }The condition (30) implies that $E_{1},E_{2},E_{0}$ are
unstable which is a necessary condition for the persistence of the system.%
\newline

\section{Conclusions}

Conditions for Turing instability and persistence in the presence of cross
diffusion are derived. The predator-prey model is generalized to the cross
diffusion case. Several types of interactions are assumed. .

Stability and persistence of the solutions of an epidemic model with cross
diffusion is studied.

The effect of cross diffusion to evolutionary games\ is studied. In some
cases, cross diffusion destabilizes the equilibrium solution.

Conditions for persistence and stability for the dynamics of learning in
multiagent systems are concluded.\newline
\newline
\textbf{Acknowledgments:}\newline

We thank R. S. Cantrell and C. Cosner for their comments. Also we thank the
referees for their helpful comments.\newline
\newline
\textbf{References}

\begin{enumerate}
\item[ ]  Ahmed E., Hegazi A. S. and Elgazzar A. S. (2002), On spatial
asymmetric{\normalsize \ }games. Advances in complex systems 5, 433-444.

\item[ ]  Cantrell R. S., Cosner C. and Hutson V. (1993), Permanence in ecological
systems with spatial heterogeneity. Proceeding of the Royal. Society of
Edinburgh 123A, 533-559.

\item[ ]  Chattopadhyay J. and Chatterjee S. (2001), Cross diffusional effects in a
Lotka-Volterra competitive system, Nonlin. Phen. in Complex Sys. 4, 364-369.

\item[ ]  Chen L., Hsiao L. and Li Y.(2003), Strong solution to a kind of cross
diffusion parabolic system, Comm. Math. Sci. 1, 799.

\item[ ]  Hofbauer J. (1999), The spatially dominant equilibrium of a game.
Annals of Operations Research 89, 233-251.

\item[ ]  Hofbauer J. and Sigmund K. (1998), Evolutionary games and
population dynamics. Cambridge university press, Cambridge.

\item[ ]  Le D. (2003), Cross diffusion systems on n-dimensional domains, Electronic
J. Diff. Eqns, Conference 10, 193.

\item[ ]  Maynard Smith J. (1982), Evolution and theory of games, Cambridge
Univ. Press U.K.

\item[ ]  Mendez V. (1998), Epidemic model with infected infectious period.
Phys. Rev. E 57, 3622-3624.

\item[ ]  Murray J.D.. (2003), ''Mathematical biology'', Springer Publishers..

\item[ ]  Okubo A. (1980), Diffusion and ecological problem.
Springer-Verlag, Berlin..

\item[ ]  Saffre F. and Deneubourg J. L. (2002), Swarming strategies for
cooperative species. J. Theor. Biol. 214, 441-451.

\item[ ]  Sato Y. and Crutchfield J. P. (2002), Coupled replicator equations
for the dynamics of learning in multiagent systems, Nlin AO/0204057.
\end{enumerate}

\end{document}